\documentclass[12pt,a4paper,notitlepage]{article}
\usepackage{ucs}                                                                                %
\usepackage[utf8x]{inputenc}                                                              %
\usepackage{textcomp}                                                                       %
\usepackage[english,french]{babel}                                                   %
\usepackage{amsmath,amssymb}                                                     %
\usepackage{pifont}                                                                              %
\usepackage{graphicx,axodraw}                                                         %
\title{Exact Renormalization Group : A New Method for Blocking the Action.}
\author{Jean-Michel Caillol\\
LPT-CNRS (UMR 8627 )\\
B\^at. 210 \\
Universit\'e de Paris-Sud \\
F-91405 Orsay Cedex\\ France}                                                                            %
\date{\today}                                                                                          
\begin{document}
\selectlanguage{english}
\maketitle
\begin{abstract}
We consider the exact renormalization group for a non-canonical scalar field theory in which the field is coupled
to the external source in a special non linear way. The Wilsonian action and the average effective  action are then simply  related
by a Legendre transformation up to a trivial quadratic form. An exact mapping between  canonical and non-canonical theories
is obtained as well as the relations between their flows. An application to the theory of liquids is sketched.
\end{abstract}
\tableofcontents
\section{Introduction}
\label{intro}
During the last twenty years Wilson approach\cite{Wilson} to the renormalization group (RG) has been the subject of a revival in
both  statistical and quantum  field theory and also, quite independently, in the more restricted domain of the equilibrium statistical physics
of classical liquids.

In field theory two main formulations of the non perturbative renormalization group (NPRG) have been developed 
in parallel. 
In the first one, a continuous realization of the RG transformation of the action $\mathcal{S}_k[\varphi]$ is made
and no expansion is involved with respect to some small parameter of this action. At scale-$k$ (in momentum space) the high energy
modes $\widetilde{\varphi}_q$, $q>k$, have been integrated out in the ``Wilsonian'' action
$\mathcal{S}_k$  which is a functional of the slow modes  $\widetilde{\varphi}_q$, $q<k$.  This operation
requires the implementation of some cut-off of the propagator aiming at separating slow ($q<k$) 
and fast ($q>k$) modes. The flow of the action is governed either by the Wilson-Polchinski equation\cite{Wilson,Wegner,Polchinski}
in case of a smooth cut-off or the  Wegner-Houghton\cite{WH} equation in case of a sharp cut-off. These equations, due to their complexity,
call for the use of approximation and/or truncation methods which have been extensively studied these last past years; 
we refer to the  review of Bagnuls and Bervillier\cite{Bervillier} for a detailed discussion of this first version of the NPRG. 

The second, more recent formulation, called the ``effective average action'' approach, was developed after the seminal works 
of Nicoll,  Chang and Stanley for 
the sharp cut-off version\cite{Ni1,Ni2} and Wetterich, Ellwanger and Morris (WEM) for the smooth cut-off 
version\cite{Wetterich00,Wetterich0,Ellwanger1,Ellwanger2,Ellwanger3,Morris}. This method
implements on the effective average action $\Gamma_k[\Phi]$ - roughly speaking the Gibbs free energy of the fast modes 
$\widetilde{\Phi}_q$, $q>k$ of the
classical field- rather than on  the Wilsonian action $\mathcal{S}_k$, the ideas of integration of high-energy modes that underlies any RG approach. 
The flow of $\Gamma_k$
 results in  equations which can be solved under the same kind of non perturbative approximations than those used for the Wilson-Polchinski or 
Wegner-Houghton equations. 
The main advantage of this more recent formulation is that it gives access to  the RG flow of physical quantities, 
i.e. the Gibbs free energy $\Gamma_k[\Phi]$ and the correlation functions as well, rather than such  a highly abstract 
object  as the Wilsonian action.  Recent reviews and lectures devoted to this second approach are
 available\cite{Wetterich,Delamotte} and should be consulted for a thorough discussion.
These two versions of the RG  are in fact equivalent; this not so obvious equivalence is discussed in details by Morris in a
beautiful paper\cite{Morris}.

As can be tracked back in the literature, the "effective average action" approach of the RG was in fact discovered independently
by Parola and Reatto in the framework of the theory
of liquids nearly 25 years ago; they  considered  both the sharp and the soft cut-off formulation of the so-called
hierarchical reference theory (HRT)\cite{paro1,paro2,Parola}; a review article resumes their early achievements\cite{Parola1}
and several papers describing new developments of the soft cut-off version of HRT appeared recently\cite{Soft1,Soft2}. 

Some years ago it was realized that a statistical field description of liquids was possible and the so-called 
KSSHE theory of liquids 
(after the names of Kac, Siegert, Stratonovich, Hubbard and Edwards\cite{Kac,Siegert,Strato,Hubbard,Edwards})
was introduced and developed   in
 references\cite{Cai-Mol,Cai-Stat1,Cai-Stat2,Cai-Stat3,Ukraine1,Ukraine2}. 
In reference\cite{Cai-Mol2} it was shown  that  the WEM equations for KSSHE field theory are identical to HRT equations in the sharp cut-off limit.
There are however differences for the soft formulations and a picture of the RG of liquids in terms of a Wilsonian action does not emerge
obviously from these early attempts.

A close inspection of KSSHE theory reveals that it is not an ``ordinary'' or ``canonical''  field theory in the sense that the coupling between
the scalar ``internal'' field and the  ``external'' source, in this very case the chemical potential, is non-linear. 
So it is slightly at variance with  the usual formulations
of field theory where a linear coupling is adopted in general. It turns out that the full RG construction is much more easy for
 a KSSHE-like theory than for a canonical one. 
Therefore ideas pertaining to the theory of liquids can be exported  to statistical field theory, yielding important simplifications for the latter.
Indeed the subtle reasonings of Morris\cite{Morris} can then be reproduced with a disarming simplicity by the introduction
of a ``reference'', non-gaussian system; by this way we find that the Wilsonian action 
$\mathcal{S}_k$ and the WEM action $\Gamma_k$ are related
by a simple Legendre transformation (up to a trivial quadratic form).
This is the main result of article.

The paper is organized as follows. In section~\ref{non-cano} we show how to build a non-canonical, KSSHE-like field theory from a canonical one.
We then follow  Morris's  construction of the RG in section~\ref{RG}, obtain the RG flows and discuss the interplay between the
Wilson-Polchinsy and WEM formulations of the renormalization group. The exact mapping onto a related canonical theory is discussed in 
section~\ref{Map}. The full set of flow equations for the effective vertices is then discussed in section~\ref{HRT} for the soft 
and sharp cut-off versions. 
Finally, in section~\ref{conclusion} we give an illustration for the theory of liquids and we conclude.

 In order  to simplify our discussions we have restricted ourselves
to the case of  bosonic scalar field theories; extensions to more complicated cases is certainly possible. Moreover we discuss here only 
but the ``first step'' of the RG program of Wilson, i.e. the blocking  of the action and we make no comments or disgressions on the scaling
properties of the solutions of  NPRG equations near a fixed point;
specialists are still at variance on this point, see e.g. reference~\cite{Bervil}.

\section{A non-canonical statistical field theory}
\label{non-cano}
To simplify the discussion let us consider a ``reference'' (R)  system described  by a standard
scalar field theory. Other representations or generalizations are easy to deal with, as illustrated in section~\ref{conclusion}
where the case of the  theory of liquids is briefly evoked.
The physics of the R-system, i.e. its thermodynamics  and correlation functions, is supposed to be known exactly;
it  is encoded in the functional \cite{Zinn, Goldenfeld}
\begin{equation} \label{Z_R}
Z_{R}\left[ J \right] = \int \mathcal{D}\chi \exp\left( -S_{R}\left[ \chi \right] 
                               + J \cdot \chi \right)  \, ,
\end{equation}
where  $J(x)$ is an external source and the action $S_{R}\left[ \chi \right]$ is an arbitrary
functional of the  real scalar field $\chi$; notably $S_{R}\left[ \chi \right]$ might comprise terms linear 
or quadratic in $ \chi$.
\mbox{In \eqref{Z_R}}  $J \cdot \chi $ is a short-cut for  $\int_{x}  J(x ) \cdot \chi(x)$ where
 $\int_{x} \equiv \int \mathrm{d}^{\mathrm{d}}x $ and $d$ the space dimensions.

We  denote by  \mbox{$W_{R}\left[ J \right] =\ln Z_{R}\left[ J \right] $} the Helmholtz free energy functional.
As well known $Z_{R}$ and $W_{R}$ are the generators of
ordinary and connected correlation functions  which will be written as
\begin{subequations}
\begin{align}
Z_{R}^{(n)}\left( J ; 1,2,\ldots,n\right) & = \dfrac{1}{Z_{R}} \dfrac{\delta^{n }Z_{R} }{ \delta J(1) \ldots\delta J(n)  }     \, ,  \\
W_{R}^{(n)}\left( J ; 1,2,\ldots,n\right)& =  \dfrac{\delta^{n }W_{R} }{ \delta J(1) \ldots\delta J(n)  }      \, ,
\end{align}
\end{subequations}
where we used the uncluttered notations  $i \equiv x_{i}$.

 From first principles $W_{R}\left[ J \right]$ is a convex functional of the source $J(x)$.
Its Legendre-Fenchel transform 
$\Gamma_{R}\left[ \Phi \right]$, the reference Gibbs free energy, is therefore also a convex
functional of the classical field $\Phi(x)$. We thus have
\begin{subequations}\label{Legendre}
\begin{align}
\Gamma_{R}\left[ \Phi \right] &= \sup_{J}\left( J \cdot\Phi - W_{R}\left[ J \right]  \right)    \\
W_{R}\left[ J \right]                &= \sup_{\Phi}\left( J \cdot\Phi - \Gamma_{R}\left[ \Phi \right]  \right)  \; ,
\end{align}
\end{subequations}
from which we deduce Young inequalities
\begin{equation}\label{Young}
\Gamma_{R}\left[ \Phi \right] +  W_{R}\left[ J \right] \geq J \cdot\Phi \qquad (\forall \Phi ,\forall J ) \, ,
\end{equation}
which may be used to obtain  rigorous bounds  (see e.g. appendix~ B).

It will  prove useful to  introduce the proper vertex  functions of the R-system  
\begin{equation}
\Gamma_{R}^{(n)}\left( \Phi ; 1,2,\ldots,n \right) =  \dfrac{\delta^{n} \Gamma_{R} }{ \delta \Phi(1) \ldots\delta \Phi(n)  }    \; .
\end{equation}

The Legendre-Fenchel transform\ \eqref{Legendre} is more general than, but in the cases that will be 
considered here equivalent
to, the usual  Legendre transform defined as
\begin{equation}
\Gamma_{R}\left[ \Phi \right] +  W_{R}\left[ J \right] = J \cdot\Phi \qquad
 \begin{cases}
\forall \Phi  &\qquad J(x)= \Gamma_{R}^{(1)}(\Phi;x ) \, , \\
\forall J      &\qquad \Phi(x)= W_{R}^{(1)}(J;x )  \, .
 \end{cases}
\end{equation}

Our requirements concerning the properties of the  R-system will be modest and fuzzy;
a reasonable assumption is that it is not at, or too close to a critical point, so that 
connected correlations functions are short ranged and Taylor functional
expansions  about some arbitrary field make sense. In practice $Z_R [J]$ is of course not known exactly 
and will in general result from some approximation, a high-temperature expansion for instance. 
The choice of a gaussian model for the R-system would obviously be either of little interest
or a lack of ambition.
At this point we introduce and want to study a family of models (referred to as $\Lambda$-systems)
labelled by $\Lambda$  in momentum space and built as follows
\begin{equation}\label{ZLambda}
Z_{\Lambda}\left[ J \right] = \int \mathcal{D}\chi \exp\left( -S_{R}\left[ \chi \right] 
+\dfrac{1}{2}\chi \cdot P^{\Lambda}_{0}\cdot  \chi  + J \cdot \chi \right)  \, ,
\end{equation}
where 
$\chi \cdot P^{\Lambda}_{0}\cdot  \chi \equiv\int_{x} \!  \int_{y}
\chi(x) P^{\Lambda}_{0}(x -y)  \chi(y) \;$,
which can also be rewritten as
$\int_{q} \widetilde{P}^{\Lambda}_{0}(q) \chi_{q}\chi_{-q} $ in Fourier space
where $\widetilde{P}^{\Lambda}_{0}$ and $\chi_{q}$ denote the Fourier transforms of
$P^{\Lambda}_{0}(x)$ and $\chi({x})$ respectively, finally
$\int_{q} \equiv \int \mathrm{d}^{\mathrm{d}}q/(2 \pi)^{d}$.
We assume that $ P^{\Lambda}_{0}$ is definite positive (i.e.  $\widetilde{P}^{\Lambda}_{0}(q) > 0$).
$\Lambda$ acts as an ultraviolet (UV) cut-off since 
\begin{subequations}\label{1}
\begin{align}
\widetilde{P} ^{\Lambda}_{0}& =       \widetilde{P}_{0}(q)  C\left( \dfrac{q}{\Lambda} \right)   \\
C(x)                            & = 1 - \Theta_{\epsilon}(x-1) \; ,
\end{align}
\end{subequations}
where $\Theta_{\epsilon}(x)$ is a smoothened version of the step function $\Theta(x)$, $\epsilon$ 
being the range of the interval $(-\epsilon/2, \epsilon/2)$ where $\Theta_{\epsilon}(x)$ increases
gently from $\Theta_{\epsilon}=0$ to $\Theta_{\epsilon}=1$.
We will denote similarly $\delta_{\epsilon}(x)=\partial \Theta_{\epsilon}(x)/ \partial x $ the smoothened version
of Dirac distribution. Taking (carefully) the limit $\epsilon \to 0$ yields  the sharp cut-off version of the theory.
In \eqref{1} $\widetilde{P}_{0}(q)  \propto 1/( q^{2} + m^{2})$ is a massive propagator, but we can find
no reason why $m^{2}$ 
could not be set to 0 if necessary.  We see  that 
$ \widetilde{P} ^{\Lambda}_{0}(q)  \approx \widetilde{P}_{0}(q) $ for $q \leq \Lambda - \epsilon$ and
$ \widetilde{P} ^{\Lambda}_{0}(q)  \approx 0 $ for $q \geq \Lambda + \epsilon$. 
The UV cut-off $\Lambda$ may be understood as the scale at which  the 
$\Lambda$-system is defined at a microscopic level; for an Ising model typically $\Lambda \approx 1/a$
 where $a$ is the lattice spacing and for a fluid
of molecules of size $\sigma$,  $\Lambda \approx 1/\sigma$. Note that since a positive quadratic term has been
\textit{substracted} to the action $S_{R}$ the $\Lambda$-system can  be tuned to  a critical point.

We now take advantage of the positivity of operator $ P^{\Lambda}_{0}$  to perform a 
 Hubbard-Stratonovich transform \cite{Kac,Siegert,Strato,Hubbard,Edwards} in~\eqref{ZLambda} which yields 
\begin{subequations}\label{strato}
\begin{align}
Z_{\Lambda}\left[ J \right] &=\frac{1}{\mathcal{N}_{P^{\Lambda}_{0}} } \,
                                             \int \! \mathcal{D}\varphi \, \exp\left(
-\dfrac{1}{2}\varphi \cdot R^{\Lambda}_{0} \cdot  \varphi         + W_{R}\left[ J + \varphi \right]  \right)  \, , \\
\mathcal{N}_{P^{\Lambda}_{0}} &=  \int  \! \mathcal{D}\varphi \, \exp\left(  
-\dfrac{1}{2}\varphi \cdot R^{\Lambda}_{0} \cdot  \varphi  
\right) \, ,
\end{align}
\end{subequations}
where $R^{\Lambda}_{0} \equiv [P^{\Lambda  }_{0}]^{-1}$ is the inverse of $P^{\Lambda}_{0}$ in the sense of operators, i.e. 
$$\int_{y} R^{\Lambda}_{0}(x,y)  P^{\Lambda}_{0}(y,z)  =
\delta^{d}(x -z) \, .$$ The Hubbard-Stratonovich transform and
other useful properties of gaussian functional integrals are  reviewed in Appendix~A.

The field theory given by \eqref{strato} is non-canonical in the sense that the coupling between the external source $J$ and the field
$\varphi$ is a non linear one.  This kind of field theory appears naturally in the  statistical mechanics of simple fluids, the Ising model, etc
after performing a Hubbard-Stratonovich transform in order to introduce a field theory for the model under consideration.
The  KSSHE theory of liquids is introduced and 
discussed  in 
references~\cite{Cai-Mol,Cai-Stat1,Cai-Stat2,Cai-Stat3,Ukraine1,Ukraine2}, some of its salient  features are reviewed in Appendix~B and 
additional comments are given in section~\ref{conclusion}.
\section{The exact renormalization group}
\label{RG}
\subsection{Blocking the action}
\label{Blocking}
We now apply the exact RG approach of Tim Morris \cite{Morris}  to our non-canonical field theory.
As a consequence of  Bogolioubov  theorem (cf.~equation~\eqref{Wick1} in Appendix~A) the partition function
$Z_{\Lambda}\left[ J \right]$ can be rewritten in terms of two propagators and two fields as
\begin{subequations}\label{Bogo}
\begin{align}
Z_{\Lambda}\left[ J \right] &=  \frac{1}{\mathcal{N}_{P^{k}_{0}} } \, 
 \int \! \mathcal{D}\varphi_{<} \, \exp\left(
-\dfrac{1}{2}\varphi_{<} \cdot R^{k }_{0} \cdot  \varphi_{<}  
\right) \, Z^{\Lambda}_{k}\left[\varphi_{<},J \right]  \, , \\
Z^{\Lambda}_{k}\left[\varphi_{<},J \right] &= \frac{1}{\mathcal{N}_{P^{\Lambda}_{k}} } \, 
 \int \! \mathcal{D}\varphi_{>} \, \exp\left(
-\dfrac{1}{2}\varphi_{>} \cdot R^{\Lambda}_{k} \cdot  \varphi_{>}  \;
+ W_{R}\left[J +\varphi_{<} + \varphi_{>} \right] 
\right) \, ,  
\end{align}
\end{subequations}
where $0\leq k\leq \Lambda$ is the running scale of the RG and where
\begin{equation}\label{decompo}
\varphi = \varphi_{<} + \varphi_{>} \text{  and  }  P^{\Lambda}_{0}=P^{\Lambda}_{k}+P^{k}_{0} \, .
\end{equation}
In \eqref{Bogo}-\eqref{decompo} we have separated the field $\varphi$ into ``rapid'' ($\varphi_{>}$)
and slow modes \mbox{($\varphi_{<}$)}. The low-energy modes are associated to the propagator
$P^{k}_{0} $ (with inverse $R^{k}_{0} $) which is cut off from above by $k$, while
 the  high-energy modes are associated to
the propagator $P^{\Lambda}_{k} $         (with inverse $R^{\Lambda}_{k} $) 
which is cut off from below by $k$ and from above by $\Lambda$. We demand  that 
$\widetilde{P}^{\Lambda}_{k}(q)=\widetilde{P}_{0}(q) (C(q/\Lambda) -C(q/k) ) $ should be positive and thus
the cut-off function $C(x)$ must be a decreasing function of its argument which will be assumed henceforth.

As in the canonical case, the functional $Z^{\Lambda}_{k}\left[\varphi_{<},J \right] $ is the crux of the whole matter
since it allows to make explicit the link between the Wilsonian action and the effective average action\cite{Morris}.
However  here this link  proves trivial since 
 $Z^{\Lambda}_{k}\left[\varphi_{<},J \right] $ is  a functional of the single variable $\varphi_{<}+J $.

Let us first set $J=0$ in \eqref{Bogo}. In the one hand we have
\begin{subequations}
 \label{Wilson}
\begin{align}
Z^{\Lambda}_{k}\left[\varphi_{<},J=0 \right]& \triangleq       \exp    \left( -S^{\Lambda}_{k}\left[\varphi_{<} \right]\right)     \label{Wilson-a} \\
&= \frac{1}{\mathcal{N}_{P^{\Lambda}_{k}} } \, 
 \int \! \mathcal{D}\varphi_{>} \,
 \exp\left(
-\dfrac{1}{2}\varphi_{>} \cdot R^{\Lambda}_{k} \cdot  \varphi_{>}  \;
+ W_{R}\left[\varphi_{<} + \varphi_{>} \right] 
\right) \,\label{Wilson-b} \, ,
\end{align}
and on the other hand
\begin{align} \label{Wilson-c}
Z_{\Lambda}\left[J=0\right]&= \frac{1}{\mathcal{N}_{P^{k}_{0}} } \, 
 \int \! \mathcal{D}\varphi_{<} \, \exp\left(
-\dfrac{1}{2}\varphi_{<} \cdot R^{k }_{0} \cdot  \varphi_{<}  - S^{\Lambda}_{k}\left[\varphi_{<}\right]
\right)   \, .
\end{align}
\end{subequations}
Equations \eqref{Wilson} define  the Wilsonian action  
$S^{\Lambda}_{k}\left[\varphi_{<} \right]$ in the usual way, i.e. as the effective action of the 
slow modes at scale $k$ \cite{Wilson,Wegner,Morris}.  Here $k$ plays the role of an UV cut-off.

Let us now set $\varphi_{<}=0$ in \eqref{Bogo}. It yields
\begin{subequations}\label{niceZ}
\begin{align}
Z^{\Lambda}_{k}\left[ \varphi_{<}=0, J \right] & \triangleq Z^{\Lambda}_{k}\left[J  \right]  (  \triangleq
    \exp    \left( W^{\Lambda}_{k}\left[J  \right]\right)   )   \label{Wett-a} \\
&= \frac{1}{\mathcal{N}_{P^{\Lambda}_{k}} } \, 
 \int \! \mathcal{D}\varphi_{>} \,
 \exp\left(
-\dfrac{1}{2}\varphi_{>} \cdot R^{\Lambda}_{k} \cdot  \varphi_{>}  \;
+ W_{R}\left[J+ \varphi_{<}  \right] 
\right) \,\label{Wett-b} \, ,
\end{align}
which shows that $W^{\Lambda}_{k}\left[J  \right]$ is the Helmholtz free energy of the rapid modes
$\varphi_{>} $ in the presence of the source $J$; therefore, here,  $k$ plays the role of an infra-red (IR) cut-off.
We will see in section \ref{map} how $W^{\Lambda}_{k}\left[J  \right]$ may also be seen, in some sense, as
the generator of connected correlation functions  with UV regularization, i.e. $\Lambda$,  and IR cut-off, i.e. $k$. 

We  also note that the partition 
function  $Z^{\Lambda}_{k}\left[J  \right]$  can  alternatively be written
 as a functional integral over the field $\chi$, i.e.
\begin{equation}\label{Wett-c}
Z^{\Lambda}_{k}\left[ J \right] = \int \mathcal{D}\chi \exp\left( -S_{R}\left[ \chi \right] 
+\dfrac{1}{2}\chi \cdot P^{\Lambda}_{k}\cdot  \chi  + J \cdot \chi \right)  \, .
\end{equation}
\end{subequations}
A Hubbard-Stratonovich transform allows indeed to obtain \eqref{Wett-b} from \eqref{Wett-c} in the same way used
to pass from the expression \eqref{ZLambda} of $Z_{\Lambda}$ to equation \eqref{strato}. 

As a trivial consequence 
of \eqref{Wett-c} we note that 
$W^{\Lambda}_{\Lambda}\left[ J \right]  =W_{R}\left[ J \right]$ (since 
\mbox{$P^{\Lambda}_{\Lambda} \equiv 0$} as follows from \eqref{decompo})
and  $W^{\Lambda}_{0}\left[ J \right]  =W_{\Lambda}\left[ J \right]$. 
Another important consequence
of $\eqref{Wett-c}$ is the convexity of the  functional $W^{\Lambda}_{k}\left[ J \right] $ 
which follows from the usual arguments \cite{Goldenfeld,Cai-conv}.

The two approaches of the RG,  that of the Wilsonian action and that of the  effective average action
are here trivially related since
\begin{equation}\label{S=W}
W^{\Lambda}_{k} = -  S^{\Lambda}_{k} \, ,
\end{equation}
from which we infer that 
\begin{equation}\label{zz}
Z^{\Lambda}_{k}\left[\varphi_{<},J \right]=\exp \left(-  S^{\Lambda}_{k} \left[\varphi_{<}+J \right] \right) 
= \exp \left( W^{\Lambda}_{k} \left[\varphi_{<}+J \right] \right) \, .
\end{equation}
When \eqref{zz} is reported in \eqref{Bogo} we get the illuminating expression
\begin{equation}\label{ille}
Z_{\Lambda}\left[ J \right] =  \frac{1}{\mathcal{N}_{P^{k}_{0}} } \, 
 \int \! \mathcal{D}\varphi_{<} \, \exp\left(
-\dfrac{1}{2}\varphi \cdot R^{k }_{0} \cdot  \varphi + W^{\Lambda}_{k}\left[J+ \varphi \right] 
\right)   \, ,
\end{equation}
which, when compared to equation~\eqref{strato}, shows that  $W^{\Lambda}_{k}$ can also be interpreted as a reference
Helmholtz free energy at scale $k$ or
as the Helmholtz free energy of the $k$-system to paraphrase Parola and Reatto \cite{Parola,Parola1}.
\subsection{Flow equations}\label{flow}
\subsubsection{The Helmholtz free energy $W^{\Lambda}_{k}$}\label{FlowW}
We first establish the flow equation for $W^{\Lambda}_{k}$.
 It follows from  expression \eqref{Wett-b} and the algebraic identity  \eqref{Wick2} of appendix~A that
\begin{equation}\label{intW}
\exp\left( W^{\Lambda}_{k}\left[ J \right]  \right) =
\exp\left(D_{k}^{\Lambda}\right) \exp\left( W_{R}\left[ J \right]  \right)  \, ,
\end{equation}
where
\begin{equation}
D_{k}^{\Lambda} \equiv \dfrac{1}{2}\int_{x,y} P_{k}^{\Lambda}(x,y)
 \dfrac{\delta}{\delta J(x)} \dfrac{\delta}{\delta J(y)} \, .
\end{equation}
Taking partial derivatives of both sides of equation \eqref{intW} with respect to $k$ at fixed $J(x)$ yields
\begin{equation}\label{flowW}
\partial_{k} \left. W^{\Lambda}_{k}\left[ J \right]  \right \vert_{J}=\frac{1}{2}
\int_{x,y} \partial_{k} P_{k}^{\Lambda}(x,y) \,
 \left\lbrace  W^{\Lambda \, (2)}_{k}\left(x ,y \right) 
+ W^{\Lambda \, (1)}_{k}\left(x \right)W^{\Lambda \, (1)}_{k}\left(y \right)
\right\rbrace \; .
\end{equation}
This flow equation must be supplemented by the initial condition $W^{\Lambda}_{\Lambda}=W_{R}$ at
$k=\Lambda$.
\subsubsection{The Wilsonian action $S^{\Lambda}_{k}$}\label{FlowS}
Since $W^{\Lambda}_{k} = -  S^{\Lambda}_{k}$ the flow of  $ S^{\Lambda}_{k}$ is given by  the usual Wilson-Polchinski
 equation \cite{Polchinski,Morris}
\begin{equation}\label{flowS}
\partial_{k} \left. S^{\Lambda}_{k}\left[  \Phi \right]  \right \vert_{\Phi}=\frac{1}{2}
\int_{x,y}\partial_{k} P_{k}^{\Lambda}(x,y) \,
 \left\lbrace  S^{\Lambda \, (2)}_{k}\left(x ,y \right) 
- S^{\Lambda \, (1)}_{k}\left(x \right)S^{\Lambda \, (1)}_{k}\left(y \right)
\right\rbrace \, ,
\end{equation}
to be supplemented with the initial condition $S^{\Lambda}_{\Lambda} = -  W_{R}$ at $k=\Lambda$. 
\subsubsection{The effective average action $\Gamma^{\Lambda}_{k}$}\label{FlowG}
The ``true'' Gibbs free energy of the $k$-system, provisionally denoted as $\overline{\Gamma}^{\Lambda}_{k}$,
 is defined as the Legendre transformation of
$ W^{\Lambda}_{k}\left[ J \right]$ and we thus have the couple of relations
 \begin{subequations}\label{Legendrek}
\begin{align}
\overline{\Gamma}^{\Lambda}_{k}\left[ \Phi \right] &= \sup_{J}\left( J \cdot\Phi - W^{\Lambda}_{k}\left[ J \right]  \right)    \, ,  \\
W_{k}^{\Lambda}\left[ J \right]                &= \sup_{\Phi}\left( J \cdot\Phi -
 \overline{\Gamma}_{k}^{\Lambda}\left[ \Phi \right]  \right)  \, ,
\end{align}
\end{subequations}
where both $W_{k}^{\Lambda}\left[ \Phi \right]$ and $ \overline{\Gamma}_{k}^{\Lambda}\left[ \Phi \right] $
are convex functionals of their arguments.
Then it follows from stationarity that
 $\partial_{k} \left. \overline{\Gamma}^{\Lambda}_{k}\left[ \Phi \right]  \right \vert_{\Phi}
=  - \partial_{k} \left. W^{\Lambda}_{k}\left[ J \right]  \right \vert_{J}$ ($\forall J$, provided that
\mbox{ $\Phi(x) = \delta W^{\Lambda}_{k}/  \delta J(x) $}  or, $\forall  \Phi$, provided that
\mbox{ $J(x) = \delta \overline{\Gamma}^{\Lambda}_{k}/  \delta \Phi(x) $}  \cite{Zinn})
from which we conclude  that
\begin{equation}\label{flowGP}
\partial_{k}\overline{\Gamma} \left. ^{\Lambda}_{k}\left[  \Phi \right]  \right \vert_{\Phi}=\frac{1}{2}
\int_{x,y}\partial_{k} P_{0}^{k}(x,y) \,
 \left\lbrace  W ^{\Lambda \, (2)}_{k}\left(x ,y \right) 
+\Phi \left(x \right) \Phi \left(y \right) 
\right\rbrace \, ,
\end{equation}
where we have used the fact that $ \partial_{k} P_{0}^{k} =-\partial_{k} P^{\Lambda}_{k} $;  we would like to
point out that  $W ^{\Lambda \, (2)}_{k}$ is the inverse of
 $\overline{\Gamma} ^{\Lambda \, (2)}_{k} =  \Gamma ^{\Lambda \, (2)}_{k} +P_{0}^{k}  $
so that \eqref{flowGP} is closed. To get rid of the non local term on the right hand side of the equation
we are led to \textit{define} the effective average action as
\begin{equation}\label{flowGdef}
\Gamma ^{\Lambda }_{k}\left[ \Phi\right]=
\overline{\Gamma} ^{\Lambda }_{k}\left[ \Phi\right] -\dfrac{1}{2}  \Phi \cdot P_{0}^{k} \cdot  \Phi \; .
\end{equation}
Note that $ \Gamma ^{\Lambda }_{k}\left[ \Phi\right]$ can be non-convex as lonk as $k>0$ since operator $P_{0}^{k}$ is definite positive.
Obviously its flow equation takes the simple form
\begin{equation}\label{flowG}
\partial_{k}\Gamma \left. ^{\Lambda}_{k}\left[  \Phi \right]  \right \vert_{\Phi}=
\frac{1}{2}
\int_{x,y}\partial_{k} P_{0}^{k}(x,y) \,
 \left\lbrace   \Gamma ^{\Lambda \, (2)}_{k} +P_{0}^{k}  \right\rbrace^{-1}
\left(x ,y \right)
 \, ,
\end{equation}
which coincides with WEM equation. This
equation must be supplemented with an initial condition. From $W_{\Lambda}^{\Lambda}=W_{R}$ it follows
that $\overline{\Gamma}^{\Lambda}_{\Lambda}= \Gamma_{R}$ and thus, from~\eqref{flowGdef}
we get\begin{equation}\label{initG}
\Gamma^{\Lambda}_{\Lambda}\left[\Phi \right] = \Gamma_{R}\left[\Phi \right] 
-\dfrac{1}{2} \Phi \cdot P_{0}^{\Lambda} \cdot \Phi .
\end{equation}

At this point some comments are in order. 
Firstly, it turns out that, as shown in appendix~B,
 the expression\ \eqref{initG} of 
$\Gamma^{\Lambda}_{\Lambda}\left[\Phi \right]$ coincides with the mean field (MF),
or tree level approximation for the  Gibbs potential 
 $\Gamma_{\Lambda}\left[\Phi \right]$, which we denote by
$\Gamma^{\Lambda}_{\textrm{MF}}\left[\Phi \right]$.
Therefore, as in the usual canonical case, the RG flow drives the effective average action
$\Gamma^{\Lambda}_{k}\left[  \Phi \right]$ from its MF value at $k=\Lambda$
to its exact value at $k=0$ by integrating fluctuations of smaller and smaller wave numbers.
Moreover it is also shown in appendix~B that 
\begin{equation}
\Gamma^{\Lambda}_{k}\left[  \Phi \right] \leq \Gamma^{\Lambda}_{\textrm{MF}}\left[\Phi \right]
 \qquad \forall \Phi(x)  \, ,
\end{equation}
i.e. $\Gamma^{\Lambda}_{\textrm{MF}}\left[\Phi \right]$ constitutes an exact upper bound for the
effective average action.

Second comment: the arguments which led us to obtain equation\ \eqref{intW} can also well be applied to 
equations\ \eqref{strato} and\ \eqref{ille} which gives
\begin{subequations}\label{semi}
\begin{align}
\exp\left( W_{\Lambda}^{}\left[ J \right]  \right)& =
\exp\left(D_{0}^{k}\right) \exp\left( W^{\Lambda}_{k}\left[ J \right]  \right) \, , \\
\exp\left( W^{\Lambda}_{k}\left[ J \right]  \right)& =
\exp\left(D_{k}^{\Lambda}\right) \exp\left( W_{R}\left[ J \right]  \right)  \, , \\
\exp\left( W_{\Lambda}^{}\left[ J \right]  \right)& =
\exp\left(D^{\Lambda}_{0}\right) \exp\left( W_{R}\left[ J \right]  \right) \, ,
\end{align}
\end{subequations}
i.e.~ the nice  semi-group law $e^{D^{\Lambda}_{0}} \ldots = e^{D^{k}_{0}} e^{D^{\Lambda}_{k}} \ldots$,
 which of course does not  trivially follow from $D^{\Lambda}_{0}= D_{0}^{k}+D_{k}^{\Lambda}$ but in addition requires 
the ``time ordering'' of operators $e^{D^{k}_{0}}$ and $e^{D^{\Lambda}_{k}}$ as well as the conditions $0 \leq k \leq \Lambda$.
\subsection{Reparametrization invariance}\label{inv}
We discuss shortly the reparametrization invariance of the theory; indeed, changing the UV cut-off from
$\Lambda$ to some  $\Lambda' \leq \Lambda$ should not change the physics at scale $k$
provided the ``new'' reference system is properly reparametrized
at scale   $\Lambda'$. We will do it for  $S$ and $\Gamma$, the two faces of our Janus RG.

Let us choose some running wave number $0 \leq k \leq\Lambda' \leq \Lambda$. 
Recall that we have $e^{-S^{\Lambda}_{k}}=e^{D_{k}^{\Lambda}} e^{-S_{R}} $
with $S_{R} =-W_{R}$ and we define
$S_{R}'=S^{\Lambda}_{\Lambda'}$.
Obviously the semi-group law $e^{D^{\Lambda}_{0}} \ldots = e^{D^{k}_{0}} e^{D^{\Lambda}_{k}} \ldots$
which was proved to be valid for $0 \leq k \leq \Lambda$ in section~\ref{FlowG} can be generalized without 
problems to the triplet  $ k \leq\Lambda' \leq \Lambda$ (the fact that the smallest wawenumber $k=0$
 in equations~\eqref{semi}  plays no role) and therefore we  have 
$e^{-S^{\Lambda}_{k}}=  e^{D_{k}^{\Lambda'}} e^{D_{\Lambda'}^{\Lambda}}e^{-S_{R}}=e^{-S^{\Lambda'}_{k}}$
with  $e^{-S^{\Lambda'}_{k}}=e^{D_{k}^{\Lambda'}} e^{-S_{R}'} $.
Therefore  $S^{\Lambda'}_{k}=S^{\Lambda}_{k}$ 
if the action of the  new reference system is indeed chosen to be $S_{R}'=S^{\Lambda}_{\Lambda'}$;
this proves the reparametrization invariance for the Wilsonian action $S^{\Lambda}_{k}$.

We turn now our attention to the effective average action $\Gamma^{\Lambda}_{k}$ 
and give two derivations of the reparametrization invariance as both are instructive.
Since $S=-W$ we have of course $W^{\Lambda'}_{k}=W^{\Lambda}_{k}$ and more generally
 $W^{\Lambda' \, (n) }_{k}=W^{\Lambda \, (n)}_{k}$. In particular the full propagators ($n=2$) 
are equal and we infer 
from the form of the flow equation\ \eqref{flowG} that
 $\partial_{k}\Gamma^{\Lambda'}_{k}[\Phi]=\partial_{k}\Gamma^{\Lambda}_{k}[\Phi]$.
It remains to examine the initial condition for $\Gamma^{\Lambda'}_{k}[\Phi]$  at $k=\Lambda'$. 
$ \Gamma^{\Lambda'}_{\Lambda'}[\Phi]$
is given by \mbox{equation\ \eqref{initG}} i.e. 
$$ \Gamma^{\Lambda'}_{\Lambda'}[\Phi] = \Gamma_{R}'\left[\Phi \right] 
-\dfrac{1}{2} \Phi \cdot P_{0}^{\Lambda'} \cdot \Phi .\, .$$
Since  $ \Gamma_{R}'\left[\Phi \right]=\overline{\Gamma}^{\Lambda}_{\Lambda'}[\Phi]$ (a direct consequence of
$S_{R}'=S^{\Lambda}_{\Lambda'}= -W^{\Lambda}_{\Lambda'}   $) it follows
from the very definition~\eqref{flowGdef}  that
$\Gamma^{\Lambda'}_{\Lambda'}[\Phi]=\Gamma^{\Lambda}_{\Lambda'}[\Phi]$. Integrating the flow
equations thus yields
$\Gamma^{\Lambda'}_{k}[\Phi]=\Gamma^{\Lambda}_{k}[\Phi]$, i.e. the 
reparametrization invariance for the effective average action. 

For a  simpler proof we start from the reparametrization invariance for the Wilsonian action. 
As $S=-W$ we
have $W^{\Lambda'}_{k}[J]=W^{\Lambda}_{k}[J]$ from which
$\overline{\Gamma}^{\Lambda'}_{k}[\Phi]=\overline{\Gamma}^{\Lambda}_{k}[\Phi]$ by
Legendre transform. Then it follows from  \mbox{equation~\eqref{flowGdef}} that 
$\Gamma^{\Lambda'}_{k}[\Phi]=\Gamma^{\Lambda}_{k}[\Phi]$; it was therefore of the utmost importance that
the quadratic form subtracted from $\overline{\Gamma} ^{\Lambda}_{k}[\Phi]$  to
define   $\Gamma ^{\Lambda}_{k}[\Phi]$ did not depend explicitely on the UV cut-off $\Lambda$.

\section{Mapping on the canonical  theory}
\label{Map}
\subsection{The mapping}
\label{map}
Commenting on equation\ \eqref{strato} we already stressed the non-canonical functional dependence of 
$Z_{\Lambda}\left[ J \right] $ upon the source $J(x)$. 
This remark holds at any scale \mbox{$0\leq k \leq \Lambda$} and
also applies to the partition function  $Z^{\Lambda}_{k}[J]$ of the $k$-system.
The $k$-independent change of variable 
\begin{equation}\label{change}
\varphi \rightarrow \varphi ^{\star}=\varphi + J
\end{equation}
 in equation\ \eqref{niceZ} obviously
allows a simple mapping on a canonical theory.  We shall 
distinguish by a superscript ``$*$''  all the quantities pertaining to this canonical theory.
Substituting  $\varphi $  for  $\varphi^{\star}$ in the expression\ \eqref{niceZ} of
$Z^{\Lambda}_{k}[J]$  readily yields
\begin{subequations}\label{mapZ}
\begin{equation}\label{mapZa}
Z^{\Lambda}_{k}[J]= e^{-\dfrac{1}{2} J \cdot R^{\Lambda}_{k} \cdot J}Z^{\Lambda \, *}_{k}[J^{*}] \, ,
\end{equation}
where $ Z^{\Lambda \, *}_{k}[J^{*}]$ reads as
\begin{equation}\label{mapZb}
Z^{\Lambda \, *}_{k}[J^{*}]= \frac{1}{\mathcal{N}_{P^{\Lambda}_{k}} } \,
                                             \int \! \mathcal{D}\varphi^{*} \, \exp\left(
-\dfrac{1}{2}\varphi^{*} \cdot R^{\Lambda}_{k} \cdot  \varphi^{*} 
        + W_{R}\left[  \varphi^{*} \right]  + J^{*}\cdot \varphi^{*} \right) \, ,
\end{equation}
and where the sources $J$ and $J^{*}$ are related by the simple linear relations
\begin{align}\label{mapZc}
J^{*}=  R^{\Lambda}_{k} \cdot J \Leftrightarrow
J=  P^{\Lambda}_{k}\cdot J^{*} \, .
\end{align}
\end{subequations}

$ Z^{\Lambda \, *}_{k}[J^{*}]$ is the standard or ``canonical'' form of the partition function
of the $k-$system. One defines as usual the Helmholtz free energy as
$W^{\Lambda \, *}_{k}=\ln Z^{\Lambda \, *}_{k}$.
The construction of the Wilsonian action  $ S^{\Lambda \, *}_{k} $ is worked out by
means of a Bogolioubov  transformation as in section~\ref{Blocking} and 
$ \Gamma^{\Lambda \, *}_{k}$ is obtained by a modified Legendre transform of
$W^{\Lambda \, *}_{k}$. Details of the derivations are to be found in the paper of Morris
(cf. \cite{Morris}). We reproduce here only his key results, rewritten however within
our notations.
First, one has (compare with equations\ \eqref{Bogo}):
\begin{subequations}\label{Bogostar}
\begin{align}
Z_{\Lambda}^{*}\left[ J^{*} \right] &=  \frac{1}{\mathcal{N}_{P^{k}_{0}} } \, 
 \int \! \mathcal{D}\varphi_{<}^{*} \, \exp\left(
-\dfrac{1}{2}\varphi_{<}^{*} \cdot R^{k }_{0} \cdot  \varphi_{<}^{*}  
\right) \, Z^{\Lambda \, *}_{k}\left[\varphi_{<}^{*},J^{*} \right]  \, , \\
Z^{\Lambda \, *}_{k}\left[\varphi_{<}^{*},J^{*} \right] &= \frac{1}{\mathcal{N}_{P^{\Lambda}_{k}} } \, 
 \int \! \mathcal{D}\varphi_{>}^{*} \, \exp\left(
-\dfrac{1}{2}\varphi_{>}^{*} \cdot R^{\Lambda}_{k} \cdot  \varphi_{>}^{*}  \;
+ W_{R}\left[\varphi_{<}^{*} + \varphi_{>}^{*} \right] \ldots \right. \nonumber \\
       &  \ldots + J^{*} \cdot   \left(   \varphi_{<}^{*} +  \varphi_{>}^{*}\right)  \bigg) \, .
\end{align}
\end{subequations}
As in section~\ref{Blocking}  the two-fields functional $Z^{\Lambda \, *}_{k}\left[\varphi_{<}^{*},J^{*} \right]$ is the key of
Janus temple. Making $J=0$ in \eqref{Bogostar} defines the Wilsonian action
\begin{align}\label{Scano}
Z^{\Lambda \, *}_{k}\left[\varphi_{<}^{*},J^{*}=0 \right] &=\exp\left( -S^{\Lambda \, *}_{k}\left[ \varphi_{<}^{*} \right]  \right) \, ,
 \nonumber \\
&=\exp \left( D^{\Lambda}_{k} \right)  \exp \left(W_{R}\left[ \varphi_{<}^{*} \right]    \right)  \, ,
\end{align}
while letting $\varphi_{<}^{*}=0$ defines the Helmholtz free energy
\begin{align}\label{Wcano}
Z^{\Lambda \, *}_{k}\left[\varphi_{<}^{*}=0,J^{*}\right] &=\exp\left( W^{\Lambda \, *}_{k}\left[ J^{*} \right]  \right) \, ,
 \nonumber \\
&=\exp \left(\dfrac{1}{2}  J^{*}\cdot P^{\Lambda}_{k} \cdot J^{*}  \right) \exp \left( D^{\Lambda}_{k} \right)  \exp \left(W_{R}\left[ J^{*} \right]    \right)  \, .
\end{align}

Below we make explicit the mapping between all star and non star quantities and compare their
RG flows.
\subsection{$W^{\Lambda}_{k}[J]$ and the Green functions}\label{WandGreen}
The canonical and non-canonical Helmholtz free energy therefore differ by a simple
quadratic form and we have,
for example  for  $W^{\Lambda \, *}_{k}$ in terms of $W^{\Lambda}_{k}$
\begin{equation}\label{mapW}
 W^{\Lambda \, *}_{k}[J^{*}] = W^{\Lambda }_{k}[J] + \frac{1}{2} J\cdot  R^{\Lambda}_{k} \cdot J
\, ,
\end{equation}
as follows either from equations\ \eqref{mapZa} or \eqref{Wcano} which therefore are thus indeed equivalent.
$W^{\Lambda \, *}_{k}[J^{*}] $ is the generator of the connected correlation functions of field  
$ \varphi^{*}$. Since $\varphi$ and $\varphi^{*}$ differ by a constant (cf \eqref{change}) their connected
correlations differ only at order $n=1$ for which $<\varphi^{*}>=<\varphi> +J $. We shall denote 
$\Phi^{*} \equiv W^{\Lambda \, * \, (1) }_{k}=< \varphi^{*}>$ the order parameter and will  adopt 
the same notation for its non-star counterpart
$\Phi \equiv W^{\Lambda \; (1) }_{k}$ although it could be misleading since, in the non-canonical case,
 $ \Phi \neq <\varphi>$. Taking the functional derivative of
both sides of \eqref{mapW} and making use of the linear relations~\eqref{mapZc} between the sources $J$ and
$J^{*}$ leads to the relations
\begin{align}\label{mapPhi}
 \Phi& = - J^{*} + R^{\Lambda }_{k} \cdot \Phi^{*} \, , \nonumber \\
 \Phi^{*}& = J + P^{\Lambda }_{k} \cdot \Phi \, .
\end{align}
By performing successive derivatives of the above relations with respect  either to $J$ or to $J^{*}$ one obtains
easily the wanted relation between the two sets of Green functions
\begin{align}\label{mapcorre}
W^{\Lambda \, * \, (2)}_{k}(1, 2)&=P^{\Lambda }_{k}(1,2) + 
                            P^{\Lambda }_{k}(1,1') P^{\Lambda }_{k}(2,2')  W^{\Lambda  \, (2)}_{k}( 1', 2') \nonumber \\
W^{\Lambda \, * \, (n)}_{k}( 1, \ldots, n)&= P^{\Lambda }_{k}(1,1') \ldots P^{\Lambda }_{k}(n,n') 
W^{\Lambda  \, (n)}_{k}(1', \ldots, n')   \textrm{ for } n \geq 3 \, .
\end{align}
where summation, i.e. space integration, over repeated indices ($n \equiv x_n$) is meant (to
unclutter notations, the functional dependence of Green functions upon the sources $J$ and $J^{*}$ was not displayed
explicitely).
\subsection{The Wilsonian action $S^{\Lambda}_{k} $}\label{SetS}
This one is easy; a serene contemplation of equations\ \eqref{Scano} and \ \eqref{Wilson} should
convince the reader that
\begin{equation}\label{mapS}
 S^{\Lambda \, *}_{k}\left[ \Psi \right]   = S^{\Lambda }_{k}\left[ \Psi \right]  \qquad ( \forall\Psi)  \, .
\end{equation}
\subsection{The effective average action  $\Gamma^{\Lambda}_{k} $}\label{GetG}
Recall that, in the canonical case, as we did in section\ \ref{RG}, one first introduces the Legendre transform $\overline{\Gamma}^{\Lambda}_{k}$
of Helmholtz free energy
\begin{equation}\label{LegStar}
\overline{\Gamma}^{\Lambda \, *}_{k}\left[ \Phi^{*}\right] +W^{\Lambda \, *}_{k}\left[J^{*} \right] =
J^{*} \cdot \Phi^{*} \qquad  \begin{cases}
\forall\Phi^{*} \qquad  J^{*}&=\delta \overline{\Gamma}^{\Lambda \, *}_{k} /  \delta \Phi^{*} \, , \\
\forall J^{*} \qquad \Phi^{*}&= \delta W^{\Lambda \, *}_{k} /  \delta J^{*}  \, .
\end {cases}
\end{equation}
Recall that $W^{\Lambda \, *}_{k}\left[J^{*} \right]$ and  $\overline{\Gamma}^{\Lambda \, *}_{k}\left[ \Phi^{*}\right]$
are both convex functionals and that the (possibly non-convex) effective average action 
$\Gamma^{\Lambda \, *}_{k} \left[ \Phi^{*}\right]$  is defined as \cite{Ellwanger1,Morris,Wetterich}
\begin{equation}
\Gamma^{\Lambda \, *}_{k}\left[ \Phi^{*}\right] =\overline{\Gamma}^{\Lambda \, *}_{k}\left[ \Phi^{*}\right]
-\dfrac{1}{2}  \Phi^{*}\cdot  R^{\Lambda}_{k}  \cdot  \Phi^{*} \, .
\end{equation}
The mapping between the non-canonical and canonical average effective actions is obtained
 from the mapping\ \eqref{mapW} between the Helmholtz free energies. A straightforward calculation yields
 \begin{subequations}\label{GtoGa}
\begin{align}
\Gamma^{\Lambda \, *}_{k}\left[ \Phi^{*}\right] &=\Gamma^{\Lambda }_{k}\left[ \Phi \right] -
 \dfrac{1}{2} \Phi \cdot    P^{\Lambda}_{0}   \cdot \Phi - \Phi \cdot\dfrac{\delta\Gamma^{\Lambda}_{k}}{\delta\Phi} \, ,
 \label{31} \\
\Phi^{*} &=  P^{\Lambda}_{0} \cdot \Phi +  \dfrac{\delta\Gamma^{\Lambda}_{k}}{\delta\Phi} \, , \label{32}
\end{align}
 \end{subequations}
or equivalently, from the ``star world'' to the ``non-star world''
\begin{subequations}\label{GtoGb}
\begin{align}
\Gamma^{\Lambda }_{k}\left[ \Phi \right]&=       
\Gamma^{\Lambda \, *}_{k}\left[ \Phi^{*}\right] - \dfrac{\delta\Gamma^{\Lambda\, *}_{k}}{\delta\Phi^{*}}\cdot \Phi^{*}
-\dfrac{1}{2} \dfrac{\delta\Gamma^{\Lambda\, *}_{k}}{\delta\Phi^{*}}
\cdot P^{\Lambda}_{0} \cdot
\dfrac{\delta\Gamma^{\Lambda\, *}_{k}}{\delta\Phi^{*}}
     \, , \\
\Phi&=  - \dfrac{\delta\Gamma^{\Lambda \, *}_{k}}{\delta\Phi^{*}} \, .
\end{align}
\end{subequations}
These expressions are quite complicated and, despite some efforts, we were unable to derive from them 
the mapping between the vertices $ \Gamma^{\Lambda \, (n)  }_{k}$ 
and $ \Gamma^{\Lambda \, * \,  (n)  }_{k}$ for a general ``n'' 
(Quite nice-looking relations are easily obtained for $n\leq 3$ but cannot be generalized in a straightforward manner for higher ``$n$'').

 An instructive consequence of equations \eqref{GtoGa}
and \eqref{GtoGb}  is the derivation of the initial condition for $\Gamma^{\Lambda \, *}_{k}\left[ \Phi^{*}\right]$. From 
the expression\ \eqref{initG} of $ \Gamma^{\Lambda}_{\Lambda}\left[\Phi \right] $
combined with equation\  \eqref{32} one gets
 $\Phi^{*}=\delta \Gamma_{R}/\delta \Phi$ ($\equiv J_{R}$ if you wish). From \eqref{31} one then infers
\begin{align}
\Gamma^{\Lambda \, *}_{\Lambda}\left[ \Phi^{*}\right] &= \Gamma_{R}\left[\Phi \right] -J_{R} \cdot \Phi \, , \nonumber \\
&=-W_{R}\left[  J_{R}\right]  \, , \nonumber  \\
&= S_{\Lambda}\left[  \Phi^{*} \right]   \, ,
\end{align}
which is indeed the expected result \cite{Ellwanger1,Morris,Wetterich}.

Our last task is to relate the flows of the effective average actions in the canonical and non-canonical
theories. The result is quite remarkable and  reads as 
 \begin{align}\label{flowandflow}
\left.\partial_{k} \Gamma^{\Lambda \, *}_{k}\left[ \Phi^{*}\right]\right \vert_{\Phi^{*}} &=
\left.\partial_{k} \Gamma^{\Lambda }_{k}\left[ \Phi\right]\right \vert_{\Phi}  \, , \nonumber \\
\textrm{ with }\Phi=  - \dfrac{\delta\Gamma^{\Lambda \, *}_{k}}{\delta\Phi^{*}} \, 
\textrm{     or      } \Phi^{*} &=  P^{\Lambda}_{0} \cdot \Phi +  \dfrac{\delta\Gamma^{\Lambda}_{k}}{\delta\Phi} \, .
 \end{align}
There are several proofs of this result; one of them being to start from equation\ \eqref{GtoGa}.
Taking its partial derivative with respect to ``k'' at fixed $\Phi^{*}$ yields
\begin{align}
\left. \partial_{k}  \Gamma^{\Lambda \, *}_{k}\left[ \Phi^{*}\right] \right \vert_{\Phi^{*}}&=
\left. \partial_{k}  \Gamma^{\Lambda }_{k}\left[ \Phi \right] \right \vert_{\Phi^{*}} +
\Phi \cdot P^{\Lambda}_{0}\cdot \left. \partial_{k}  \Phi \right \vert_{\Phi^{*}} -
\Phi^{*} \cdot \left. \partial_{k}  \Phi \right \vert_{\Phi^{*}}   \, , \nonumber \\
&=\left. \partial_{k}  \Gamma^{\Lambda }_{k} \left[ \Phi \right] \right\vert_{\Phi} + 
\left.\partial_{k}  \Phi \right \vert_{\Phi^{*}}\cdot
 \left\lbrace
\dfrac{\delta\Gamma^{\Lambda }_{k}}{\delta\Phi} +  P^{\Lambda}_{0}\cdot\Phi
-\Phi^{*}
 \right\rbrace       \, , \nonumber \\
&=\left. \partial_{k}  \Gamma^{\Lambda }_{k} \left[ \Phi \right] \right \vert_{\Phi}    \textrm{ QED }  \, ,
\end{align}
where we made use of\ \eqref{32} to obtain the last line.

A second, more direct proof of equation\ \eqref{flowandflow}  gives us the opportunity
 to write the well-known WEM equation for  $\Gamma^{\Lambda \, *}_{k}$
which we present with simplified notations as
\begin{equation}\label{flowGPcano}
 \left.\partial_{k} \Gamma^{\Lambda\, *}_{k}\left[  \Phi^{*} \right]  \right \vert_{\Phi^{*}}=\frac{1}{2}
\partial_{k} R^{\Lambda}_{k}(1,2) \,
 W ^{\Lambda \, * \, (2)}_{k}\left(1 ,2\right) +
\partial_{k} \ln \mathcal{N}_{P^{\Lambda}_{k}}
 \, .
\end{equation}
The second contribution to the r.h.s. of \eqref{flowGPcano} involves the normalization 
$\mathcal{N}_{P^{\Lambda}_{k}} $; It is independent of the field and for that reason generally not mentionned in the literature, 
here, however we need it to complete our proof. Clearly
\begin{align}
\partial_{k}\ln \mathcal{N}_{P^{\Lambda}_{k}}&=-\dfrac{1}{2} \left\langle 
\varphi(1) \varphi(2) \right\rangle_{P^{\Lambda}_{k}} \partial_{k} R^{\Lambda}_{k}(1,2) \nonumber \\
&=-\dfrac{1}{2} P^{\Lambda}_{k}(1,2) \partial_{k} R^{\Lambda}_{k}(1,2) \, ,
\end{align}
where the brackets denote a Gaussian average (see appendix A) and we made use of Wick's theorem.
To go further we remark that $\partial_{k} R^{\Lambda}_{k}=
- R^{\Lambda}_{k}\cdot   \partial_{k} P^{\Lambda}_{k} \cdot      R^{\Lambda}_{k} $ and also 
make use of the relations between canonical and non-canonical Green functions (cf equations\ \eqref{mapcorre}). 
This gives us
\begin{align}
 \left.\partial_{k} \Gamma ^{\Lambda\, *}_{k}\left[  \Phi^{*} \right]  \right \vert_{\Phi^{*}}&=
-\dfrac{1}{2} \partial_{k} P^{\Lambda}_{k}(1,2)
 \left\lbrace  W^{\Lambda  (2)}_{k}\left(1 ,2\right) + R^{\Lambda}_{k}(1,2) \right\rbrace  \ldots \nonumber \\
& \ldots -\dfrac{1}{2} P^{\Lambda}_{k}(1,2) \partial_{k} R^{\Lambda}_{k}(1,2)  \nonumber \\
&= -\dfrac{1}{2} \partial_{k} P^{\Lambda}_{k}(1,2)  W^{\Lambda  (2)}_{k}\left(1 ,2\right)     \nonumber \\
&=\dfrac{1}{2} \partial_{k} P^{k}_{0}(1,2)  W^{\Lambda  (2)}_{k}\left(1 ,2\right)
 \, ,
\end{align}
which indeed is equal to $\left. \partial_{k}  \Gamma^{\Lambda }_{k} \left[ \Phi \right] \right \vert_{\Phi}$ 
(cf equation\ \eqref{flowG}).
\section{The RG hierarchy}
\label{HRT}
\subsection{Smooth cut-off}\label{smooth}
In this section we comment on the flow equation\ \eqref{flowG} for the average effective action of our non-canonical
KSSHE-like field theory.
Henceforth we consider only  homogeneous systems, then, as usual, in momentum space, we factor out and evaluate
 the momentum conserving $\delta$ function so $n$-point correlation functions  
$\widetilde{\Gamma}^{\Lambda \, (n) }_{k}(q_1,\ldots,q_n)$ are defined only when $q_1 + \ldots +q_n=0$.
 More precisely one has for instance
\begin{align}
\widetilde{\Gamma}^{\Lambda \, (n) }_{k}(q_1,\ldots,q_n)& = \widehat{\delta}(q_1+\ldots q_n)
\int_{x_1 \ldots x_n}\exp\left( i\left( q_1 x_1 +\ldots q_{n-1}x_{n-1}\right)\right) \times  \nonumber \\
&\times \Gamma^{\Lambda \, (n) }_{k}        (x_1,\ldots,x_{n-1},0) \; , 
\end{align}
where $\widehat{\delta}(q) \triangleq (2\pi)^{\mathrm{d}}\delta^{\mathrm{d}}(q)$.
In addition, in two-point functions, we solve $q_1=-q_2=q$ and recognize that they are functions only of
 $q^{2}$ and we write them $\widetilde{\Gamma}^{\Lambda \, (2) }_{k}(q^{2})$. In the same vein we denote the
 Fourier transform of the full propagator at scale ``$k$'' 
$\widetilde{W}^{\Lambda \, (2) }_{k}(q^{2})\equiv 
1/(\widetilde{\Gamma}^{\Lambda \, (2) }_{k}(q^{2}) + P^{k}_{0}(q^{2}))$. For a uniform background field $\Phi$
equation\ \eqref{flowG} can thus be rewritten as:
\begin{equation}\label{flowA-final}
 \partial_{k} U^{\Lambda}_{k} \left[  \Phi \right]   =
\dfrac{1}{2} \int_{q}  \dfrac{\partial_{k} P^{k}_{0}(q^{2})}{\widetilde{\Gamma}^{\Lambda \, (2) }_{k}[\Phi;q^{2}]+ P^{k}_{0}(q^{2})} \, .
\end{equation}
where we have introduced the potential  $ U^{\Lambda}_{k}= \Gamma^{\Lambda}_{k}/V$, where $V$ is the volume.

To paraphrase B. Delamotte \cite{Delamotte} this beautiful equation  is exact and thus horribly complicated. Mathematically
it is a functional parabolic partial derivative equation since both $\Gamma^{\Lambda}_{k} \left[  \Phi \right]  $ and
$\widetilde{\Gamma}^{\Lambda \, (2) }_{k} [\Phi;q^{2}]$ are functionals of $\Phi$. 
As  for canonical theories \cite{Wetterich} one can, by functional derivation with respect to the field,
deduce from equation\ (\ref{flowA-final}) an infinite hierarchy of equations for 
the effective vertices $\widetilde{\Gamma}^{\Lambda \, (n) }_{k}(q_1,\ldots,q_n)$.  These equations
are better represented graphically with the help of 
Feynman diagrams. The latter will be build from the vertices
 \begin{subequations}
 \begin{align}
\begin{picture}(55,35)(0,7)
\Vertex(20,10){3}
\ArrowLine(5,20)(20,10)
\ArrowLine(5,0)(20,10)
\ArrowLine(35,20)(20,10)
\ArrowLine(35,0)(20,10)
\Text(45,20)[]{$q_1$}
\Text(-5,20)[]{$q_2$}
\Text(45,0)[]{$q_n$}
\Text(-5,0)[]{\ldots}
\end{picture}& = \widetilde{\Gamma}^{\Lambda \, (n) }_{k}(q_1,\ldots,q_n)\; \;  \; \; (n \geq 2) \; \; ,
\end{align}
the propagator 
\begin{equation}
\begin{picture}(55,35)(0,7)
\ArrowLine(0,10)(23,10)
\ArrowLine(46,10)(23,10)
\Text(10,20)[]{$q$} \Text(30,20)[]{$-q$}\end{picture} 
= \; \;\widetilde{W}^{\Lambda \, (2) }_{k}(q^{2}) \; ,
\end{equation}
and the insertion
\begin{align}
\begin{picture}(55,35)(0,7)
\ArrowLine(0,10)(20,10)
\BBoxc(23,10)(6,6)
\ArrowLine(46,10)(26,10)
\Text(10,0)[]{$q$}  \Text(36,0)[]{$-q$}
\end{picture} 
&= \; \; \partial_k \widetilde{P}^{k}_{0}(q^{2}) \, . \\  \nonumber
\end{align}
 \end{subequations}
For instance\ \eqref{flowA-final} takes the form
\begin{align}
\label{flowf-diag}\partial_k  U^{\Lambda}_{k}  & = \frac{1}{2}
\begin{picture}(55,40)(0,7)
\ArrowArc(20,10)(15,0,180)
\ArrowArcn(20,10)(15,360,180)
\Text(20,32)[]{$q$} \Text(20,-12)[]{$-q$} 
\BBoxc(5,10)(5,5)
\end{picture} \;  \; .\\
\nonumber
\end{align}
Since 
\begin{subequations}\label{grrules}
\begin{equation}
(2 \pi)^{\textrm{d}} \frac{\delta \widetilde{\Gamma}^{\Lambda \, (n) }_{k}(q_1,\ldots,q_n)}{\delta \widetilde{\Phi}_{-q}}=
\widetilde{\Gamma}^{\Lambda \, (n+1) }_{k}(q_1,\ldots,q_n,q) \, ,
\end{equation}
applying the  functional $\delta/{\delta \widetilde{\Phi}_{-q}}$ on a vertex with ``n'' legs gives rise to a vertex with ``n+1'' legs
while,  on a propagator,  this operation creates a vextex with 3 legs since
\begin{equation}
(2 \pi)^{\textrm{d}} \frac{\delta \widetilde{W}^{\Lambda \, (2) }_{k}(q_1,q_2)}{\delta \widetilde{\Phi}_{-q}}=
-\widetilde{W}^{\Lambda \, (2) }_{k}(q_1,-r_1) \widetilde{\Gamma}^{\Lambda \, (3) }_{k}(r_1,r_2,q_2) 
\widetilde{W}^{\Lambda \, (2) }_{k}(-r_2,-q_2) \, .
\end{equation}
\end{subequations}
With these rules in mind one easily obtains the first equations of the hierarchy 
\begin{align}
\label{flow-C1k}
\partial_k \widetilde{\Gamma}^{\Lambda \, ,(1)}_{k}(0) &=
\partial_k\begin{picture}(65,45)(0,7)
\ArrowLine(15,10)(55,10)\Vertex(55,10){3} \Text(35,20)[]{$0$}
\end{picture} 
= \; \; \frac{1}{2}\; \; \begin{picture}(55,45)(0,7)
\ArrowLine(0,10)(30,10) \Text(15,20)[]{$0$}
\Vertex(30,10){3}	   
\ArrowArc(45,10)(15,90,180) \Text(35,30)[]{$q$}
\ArrowArcn(45,10)(15,270,180) \Text(35,-10)[]{-$q$}
\ArrowArcn(45,10)(15,90,0)   \Text(55,30)[]{-$q$}
\ArrowArc(45,10)(15,270,360)  \Text(55,-10)[]{$q$}
\BBoxc(60,10)(5,5) 
\end{picture} \nonumber \\ \nonumber \\
&=- \frac{1}{2}\int_q\widetilde{\Gamma}^{\Lambda \, (3)}_{k}(0,q,-q) \,
\widetilde{W}^{\Lambda \, (2) }_{k}(q^{2}) \, \partial_k \widetilde{P}^{k}_0(q^{2}) \; ,
\end{align}
and
\begin{align}
\label{flow-C2k}\partial_k \widetilde{\Gamma}^{\Lambda \, (2)}_{k} (p^{2}) &=
\partial_k\begin{picture}(65,45)(0,7)
\ArrowLine(0,10)(30,10) \Text(15,20)[]{$p$}
\Vertex(30,10){3}
\ArrowLine(60,10)(30,10) \Text(45,20)[]{$-p$}
\end{picture}  \nonumber \\
&=-\frac{1}{2} \; \; \begin{picture}(65,45)(0,7)
\ArrowLine(0,0)(30,10) \Text(15,-5)[]{$p$}
\Vertex(30,10){3}
\ArrowLine(60,0)(30,10) \Text(45,-5)[]{$-p$}
\ArrowArcn(30,25)(15,180,90)
\ArrowArc(30,25)(15,0,90)
\ArrowArcn(30,25)(15,360,270)
\ArrowArc(30,25)(15,180,270)
\Text(50,35)[]{$q$}
\Text(50,15)[]{$-q$}
\Text(5,35)[]{$-q$}
\Text(10,15)[]{$q$}
\BBoxc(30,40)(5,5)
\end{picture}\;  +\; \; \; \; \; 
\begin{picture}(110,45)(0,7)
\ArrowLine(0,10)(30,10) \Text(-5,15)[]{$p$} \Text(25,20)[]{$q$}
\Vertex(30,10){3}
\Vertex(60,10){3}
\ArrowLine(90,10)(60,10)\Text(100,15)[]{$-p$} \Text(65,20)[]{$-q$} 
\ArrowArc(45,10)(15,45,90)
\ArrowArc(45,10)(15,135,180)
\ArrowArc(45,10)(15,270,360) \Text(62,-10)[]{$p+q$}
\ArrowArcn(45,10)(15,270,180)\Text(22,-10)[]{$-p-q$}
\ArrowArcn(45,10)(15,135,90)
\ArrowArcn(45,10)(15,45,0)
\BBoxc(45,25)(5,5) \Text(40,37)[]{$-q \; \; \; q$}
\end{picture} \nonumber \\
\nonumber \\
\nonumber \\
&=-\frac{1}{2}\int_q \widetilde{\Gamma}^{\Lambda \, (4) }_{k}(p,q,-q,-p)
\widetilde{W}_{k}^{\Lambda \, (2) }(q)^2 \partial_k \widetilde{P^{k}_{0}}(q^{2})  \nonumber \\
&+ \int_q \widetilde{\Gamma}^{\Lambda \, (3) }_{k}(p,q,-p-q) \widetilde{\Gamma}^{\Lambda \, (3) }_{k}(-q,-p,p+q)
\times   \nonumber \\
& \times
\widetilde{W}_{k}^{\Lambda \, (2)}(q)^2 \widetilde{W}_{k}^{\Lambda \, (2)}(p+q)
 \partial_k  \widetilde{P^{k}_{0}}(q^{2}) \; .
\end{align}
and so on. These tower of equations has exactly the same structure for the canonical and non-canonical theories
with the replacement  $R^{\Lambda}_{k} \rightarrow P^{k}_{0}$.
Flow equations for $\widetilde{\Gamma}_k^{\Lambda \, (n)}$ of higher orders are obtained in the same vein 
by making use \textit{ad libitum} of the diagrammatic rules which are   deduced from equations\ (\ref{grrules}).
Some comments are in order.
\begin{itemize}
\item The equation for $\partial_k \widetilde{\Gamma}_k^{\Lambda \, (n)}$ involves \textit{inter alias} the proper vertex
$\partial_k \widetilde{\Gamma}_k^{\Lambda \, (n+1)}$ and  
$\partial_k \widetilde{\Gamma}_k^{\Lambda \, (n+2)}$, 
therefore the hierarchy never closes. Possible approximations consist in enforcing 
a closure at some order $n$\cite{Wetterich,Delamotte,Parola1}.
\item  A little thought reveal that the one-loop structure is present at each order $n$ of the hierarchy
and therefore only one  integral on internal  variables survives. 
\item All the expressions  for the odd $\partial_k \widetilde{\Gamma}_k^{\Lambda \, (2n+1)}$ include
 diagrams with at least one odd vertex $\widetilde{\Gamma}_k^{\Lambda \, (2m+1)}$  ($m \leq n$).
Therefore, if at some scale $k$
all the odd  $\widetilde{\Gamma}_k^{\Lambda \, (2n+1)}$
happen to vanish they will remain exactly zero at smaller scales k. 
\end{itemize}
\subsection{Sharp cut-off}\label{sharp}
To extract the limit $\epsilon \to 0$ of the flow equations 
 one makes use of the "little lemma" of Morris \cite{Morris} which states that, for $\epsilon \to 0$ 
\begin{equation}
\label{lemme}
\delta_{\epsilon}(q,k)f( \Theta_{\epsilon}(q,k),k) \to
\delta(q-k) \; \int_{0}^{1}dt \; f(t,q) \;,
\end{equation}
provided that the function $f( \Theta_{\epsilon}(q,k),k)$ is continuous at $k=q$ in the limit $\epsilon \to 0$, which is the case here.
Applying lemma\ (\ref{lemme}) to equation~\eqref{flowA-final} one obtains the flow of the potential
\begin{equation}\label{yy}
 \partial_{k} U^{\Lambda}_{k} \left[  \Phi \right]   =\dfrac{1}{2} k^{d-1} \dfrac{S_{d}}{(2 \pi)^{d}}
\ln \left(1 +           \dfrac{ P_{0}(q^{2})}{{\Gamma}^{\Lambda \, (2) }_{k}\left[ \Phi;q^{2}\right] } \right)  \, ,
\end{equation}
where $S_{d}=2\pi^{d/2}\Gamma(d/2)$ if the surface of the d-dimensional sphere of radius ``1''.
The flow equations for the proper vertices $\widetilde{\Gamma}_k^{\Lambda \, (n)}$ of order $n\geq 1$   can also be obtained 
in the sharp cut-off limit
from those of previous section\ \ref{smooth} by applying the "little lemma". One finds that these equations are 
identical to those  obtained for the first time nearly 25 years ago by Parola and Reatto in the context of the theory of liquids \cite{paro1,paro2}.
Equation~\eqref{yy} is still older and  was obtained in the early ages of the RG. \cite{WH,Ni1,Ni2}
\section{Conclusion}\label{conclusion}
The main result of this paper is contained in equation~\eqref{S=W} which states that in a non-canonical, KSSHE-like field theory the Wilsonian
action $ S^{\Lambda}_{k} $  of the renormalization group coincides with the Helmholtz free energy of the k-system. 
The average effective  action $ \Gamma^{\Lambda}_{k} $  can
thus be obtained as a  Legendre transform of $ S^{\Lambda}_{k} $ (up to a trivial quadratic form).  We have derived the RG flow equations form
 $ S^{\Lambda}_{k} $ and  $ \Gamma^{\Lambda}_{k} $ 
and proved some important properties such that parametrization invariance. The exact mapping of section~\ref{Map} 
which relates the non-canonical and canonical theories show interesting features and can also be seen as a practical method
to build a KSSHE-like theory from a standard one.

As an illustration let us consider the theory of liquids.
Let the fluid be made  of identical hard spheres (HS) of diameter $\sigma$ with additional isotropic pair interactions
 $v(r_{ij})$ ($r_{ij}=| x_i -x_j |$, $x_i$  position of particle``$i$"). Since $v(r)$ is an arbitrary function of $r$  in the core,
 i.e. for $r \leq \sigma$, one can assume that $v(r)$ has been regularized in the core in such a way that  its Fourier transform $\widetilde{v}_{q}$
 is a well behaved function of $q$ and that $v(0)$ is a finite quantity. 
We denote by $\Omega$ the domain occupied by the molecules of the fluid.
For convenience $\Omega$ is supposed to be a cube of side $L$ and periodic boundary (PB) conditions 
are imposed so that the volume of $\Omega$ is $V=L^d$.
The fluid is at equilibrium in the grand canonical (GC) ensemble, $\beta=1/k_{\boldsymbol{B}}T$ is the inverse temperature
($k_{\mathrm{B}}$ Boltzmann's constant), and $\mu$  the chemical potential. In addition the particles are subject to an external
 potential $\psi(x)$ and we will denote by $\nu(x)=\beta (\mu-\psi(x))$ the dimensionless local chemical potential. 
We stick to notations usually adopted in standard textbooks devoted to the theory of liquids (see e.g. \cite{Hansen}) 
and  thus denote by $w_{0}(r)=-\beta v(r)$  \emph{minus } the dimensionless pair interaction.
Moreover we  restrict ourselves to the case of attractive interactions, i.e. such that $\widetilde{w}_{0}(q)>0$ for all $q$.

In a given GC configuration $\mathcal{C}\equiv(N;x_1 \ldots x_N)$ of the grand canonical ensemble 
the microscopic density of particles at point $x$  reads
$\widehat{\rho}(x|\mathcal{C}) = \sum_{i=1}^{N} \delta^{d}(x-x_i)$ 
and the grand canonical partition function (GCPF)  $\Xi\left[ \nu \right] $ which encodes all the physics of the model at
 equilibrium is defined  as \cite{Hansen}
\begin{align}
\label{csi}\Xi\left[ \nu \right] &=
\mathrm{Tr}\left[ \; \exp\left( -\beta \mathcal{H}_{\mathrm{GC}}
\right) \right] \; , \nonumber \\
-\beta \mathcal{H}_{\mathrm{GC}}&= -\beta
V_{\mathrm{HS}}[\mathcal{C}]+\frac{1}{2} 
\widehat{\rho} \cdot  w_{0} \cdot \widehat{\rho}  +
 \overline{\nu} \cdot \widehat{\rho}  \nonumber  \; ,\\ 
 \mathrm{Tr}\left[  \ldots \right] &=
 \sum_{N=0}^{\infty}
\frac{1}{N!} \int_{\Omega}d1 \ldots dn \ldots \; ,
\end{align}
where $i \equiv x_i $ and $di\equiv d^{d}x_i$. 
In equation\ (\ref{csi}) $\beta V_{\mathrm{HS}}\left[ \mathcal{C}\right] $ denotes 
the HS contribution to the configurational energy (i.e. $+\infty$ if there is an overlap of spheres, $0$ otherwise) 
and $\overline{\nu}=\nu+\nu_S$ where $\nu_S= - w_{0}(0)/2$ is $\beta$ times   the  self-energy of a
particle.
For a given volume $V$ and a given inverse temperature
 $\beta$, $\Xi\left[ \nu \right]$ is  a log-convex functional of the local chemical potential $\nu(x)$  \cite{Goldenfeld,Cai-conv} 

We now perform a Hubbard-Stratonovich transform to get the KSSHE representation\cite{Cai-Mol}
\begin{align}\label{blingbling}
\Xi\left[ \nu \right] &=
\mathcal{N}_{w_{0}^{\Lambda}}^{-1} \int \mathcal{D} \varphi \;
\exp \left( -\frac{1}{2} \varphi \cdot  w_{0}^{\Lambda \, -1}  \varphi 
+ \ln \Xi_{\text{HS}}\left[ \nu - \dfrac{1}{2} w_{0}^{\Lambda}(0)  + \varphi \right]
 \right) \, ,
\end{align}
where $\widetilde{w}_{0}^{\Lambda}(q)=C(q/\Lambda) \widetilde{w}_{0}(q) $ and $ \mathcal{D} \varphi $ is
Wegner's measure (cf equation~\eqref{dphi} of Appendix~A). We stress that $C(x)$ is the same UV 
cut-off function we met 
in section~\ref{non-cano};  we have used the fact that for $\Lambda \sim 1/\sigma$,  $w_{0}(r)$ and  
$w_{0}^{\Lambda}(r)$ differ only but inside the core. In
 equation~\eqref{blingbling} $\Xi_{\text{HS}}\left[ \nu -  w_{0}^{\Lambda}(0)/2  + \varphi \right]$ denotes of course the
GCPF of bare hard spheres subject to the local chemical potential $ \nu -  w_{0}^{\Lambda}(0)/2  + \varphi$.
   Comparing~\eqref{blingbling} with equation~\eqref{strato} we note the one to one correspondence
$W_{R} \longleftrightarrow  \ln \Xi_{\text{HS}}$ and $w \longleftrightarrow P$. Pair potentials correspond to propagators
and $W_R$ is the grand-potential of the HS fluid. 
Note that massive propagators in field theory correspond to attractive Yuhawa pair potentials in liquid theory.
The RG construction detailed in the core of the paper can be redone (the slight
modification due to the introduction of the self-energy  $w_{0}^{\Lambda} (0)$ in equation~\eqref{blingbling} does not spoil the result).
The k-system can thus be identified with a fluid of hard spheres interacting through the pair potentials
\begin{equation}
\widetilde{w}_{k}^{\Lambda}(q) =   \left( C(q/\Lambda) -C(q/k) \right) \widetilde{w}_{0}(q) \, .
\end{equation}
In direct space $w_{k}^{\Lambda}(r)$ is a short range potential equal to $w_{0}(r)$ for $1/\Lambda \equiv\sigma <r<1/k$ and equal to 0 for
$r>1/k$, precisely the kind of potential used in numerical simulations involving boxes of side $L=1/k$.   This supports a real space RG 
interpretation where, at scale ``k'',  $W_{k}^{\Lambda}$ is the Helmholtz free energy of a ``block'' of size $1/k$.

Generalizations to repulsive (including Coulomb interactions for instance) or  even not definite pair potentials are  possible,  a
detailed  analysis  will be given elsewhere. Of course this sketchy discussion of the KSSHE theory for a liquid could also
be extended  in the same vein and with identical conclusions to many other models of condensed matter physics such as the lattice gas
or the Ising model.

\section*{Acknowledgments}
The author acknowledges  U. Ellwanger for an interesting discussion, O. Patsahan
and I. Mryglod for useful comments and C. Bervillier for an exchange of e-mails.
\appendix
\label{appendixA}
\section{Gaussian measures and integrals}
In this appendix we give some properties on Gaussian integrals used in the main text.
Let us consider a real scalar field $\varphi(x)$ defined in a cube 
$\mathcal{C}_d$ of side $L$ and volume $V=L^d$. We assume periodic boundary conditions, i.e. 
we restrict ourselves to fields which can be expressed as a Fourier series,
\begin{equation}
\varphi(x)=\frac{1}{V} \; \sum_{q \in \Lambda}
\widetilde{\varphi}_{q} \; e^{i q \cdot x } \; ,
\end{equation} 
where $\Lambda = (2 \pi/L)\;  \mathbb{Z}^\mathrm{d}$ is the reciprocal cubic lattice ($\mathbb{Z}$ set of integers).
The reality of $\varphi$ implies that, for $q\ne 0$ 
$\widetilde{\varphi}_{-q} = \widetilde{\varphi}_{q}^{\star}$, where the star means complex conjugation.
Following Wegner \cite{Wegner} we define the normalized functional measure 
$\mathcal{D}\varphi$ as
\begin{subequations}
\label{dphi}
\begin{align}
\mathcal{D} \varphi & \equiv  \prod_{q \in \Lambda}
\frac{d \widetilde{\varphi}_{q} }
{\sqrt{2 \pi  V}} \\
d \widetilde{\varphi}_{q} d \widetilde{\varphi}_{-q} & =  2 \; d\Re{\widetilde{\varphi}_{q}} \; d\Im{\widetilde{\varphi}_{q}} 
\text{ for } q \ne 0 \, .
\end{align}
\end{subequations}
Equation\ (\ref{dphi}) can  be conveniently rewritten as 
\begin{equation}
\label{dphi_bis}
\mathcal{D} \varphi=
\frac{d \varphi_{0}}{\sqrt{2 \pi  V}}
\prod_{q \in \Lambda^{\star}}
\frac{d\Re{\widetilde{\varphi}_{q}} \; d\Im{\widetilde{\varphi}_{q}} }{\pi V} \; ,
\end{equation} 
where the sum in the r.h.s runs over only the half $\Lambda^{*}$ of
all the vectors of the reciprocal lattice
$\Lambda$ (for instance those with $q_x \geq 0$). With these definitions
one has
\begin{align}\label{norma}
\mathcal{N}_{w}&\equiv  \int \! \mathcal{D} \varphi \;  \exp \left( -\frac{1}{2}
 \varphi \cdot w^{-1} \cdot \varphi  \right)  \, , \nonumber \\
&= \exp \left(\frac{1}{2} \sum_{q \in \Lambda}  \ln \widetilde{w}(q) \right)  
\xrightarrow{L \to \infty}  \exp \left( \frac{V}{2}\;
\int_q \ln \widetilde{w}(q) \right) \; ,
\end{align}
where $w$ is definite and positive.

We define the Gaussian measure $ \mathrm{d} \mu_{w}\left[ \varphi \right] = \mathcal{N}_{w}^{-1} \, \mathcal{D} \varphi$
and the Gaussian average $\left\langle \mathcal{F}\left[ \varphi \right] \right\rangle_{w}
 =\int\mathrm{d} \mu_{w}\left[ \varphi \right]  \mathcal{F}\left[ \varphi \right] $ and recall the well known
Wick's theorem
\begin{equation}
\left\langle  \varphi(x_{1}) \ldots  \varphi(x_{n})  \right\rangle_{w} = \left\lbrace 
\begin{array}{cl}
0 &  \text{if  $n$ odd}    \, , \\
\displaystyle{\sum_{\mathrm{pairs}}} w(x_{i_{1}},
 x_{i_{2}}) \ldots w(x_{i_{n-1}}, x_{i_{n}}) &  \text{if  $n$ even}  \, .
\end{array} \right.      \label{Wick}
\end{equation}
From Wick's theorem one deduces the important result
\begin{subequations}
\begin{align}\label{Wick0}
\left\langle \exp\left( J \cdot \varphi \right)  \right\rangle_{w} &= \exp\left(\dfrac{1 }{2} J \cdot w \cdot     J  \right)  \, , \\
\left\langle \exp\left( i  J \cdot \varphi \right)  \right\rangle_{w} &=\exp\left(- \dfrac{1 }{2} J\cdot w \cdot     J \right)  \, ,
\end{align}
\end{subequations}
where $J(x)$ is a  real scalar field. Another consequence of Wick's theorem \eqref{Wick} is the following identity
 involving $n$ Gaussian measures  $ \mathrm{d} \mu_{w}^{i}\left[ \varphi_{i} \right]$, $i=1,\ldots,n$,  which is sometimes
 referred to as the Bogolioubov theorem:
\begin{equation}\label{Wick1}
\int\mathrm{d} \mu_{w_{1}+ \ldots +w_{n} }\left[ \varphi \right] \,   \mathcal{F}\left[ \varphi \right]=
\int \prod_{i=1}^{n} \mathrm{d} \mu_{w_{i}}\left[ \varphi_{i} \right]  \,  \mathcal{F}\left[ \varphi_{1}+ \ldots+ \varphi_{n} \right] \, ,
\end{equation}
where $\mathcal{F}\left[ \varphi \right]$ is some arbitrary functional of the field $\varphi$.

The last formal  consequence of Wick's theorem that we need mention is
\begin{subequations}
\begin{equation}\label{Wick2}
\int\mathrm{d} \mu_{w }\left[ \varphi \right] \,   \mathcal{F}\left[ \varphi + \varphi_{0} \right]=
\exp\left( D \right) \mathcal{F}\left[ \varphi_{0} \right]   \, ,
\end{equation}
where the functional Laplacian operator $D$ is defined as
\begin{equation}
D \equiv \dfrac{1}{2}\int_{x,y} w(x,y)
 \dfrac{\delta}{\delta\varphi(x)} \dfrac{\delta}{\delta\varphi(y)} \, .
\end{equation}
\end{subequations}
\label{appendixB}
\section{KSSHE theory}
We review some properties of a system described by a non-canonical KSSHE partition
function
\begin{subequations}
\begin{align}
Z_{\Lambda}\left[ J \right] &=\frac{1}{\mathcal{N}_{P^{\Lambda}_{0}} } \,
                                            \int \! \mathcal{D}\varphi \,
 \exp\left(  -  \mathcal{H}_{J}\left[ J,\varphi\right] \right)  \, , \\
\mathcal{H}_{J}\left[ J,\varphi\right] &=  \dfrac{1}{2}\varphi \cdot R^{\Lambda}_{0} \cdot  \varphi 
 - W_{R}\left[ J + \varphi \right] \,  ,
\end{align}
\end{subequations}
more details will be found in references \cite{Cai-Mol} and \cite{Cai-Mol2}.
In fact, we have already studied the Green functions of the model  in section\ \ref{WandGreen} since 
$Z_{\Lambda}\left[ J \right]$ is nothing but the special case $Z^{\Lambda}_{k=0}\left[ J \right]$.
In particular $$\Phi_{\Lambda}[J;1]\equiv W^{\Lambda \; (n=1)}_{0}(J; 1)=R^{\Lambda}_{0}(1,2) \cdot <\varphi(2)>$$
and the correlations of higher order are given by equations\ \eqref{mapcorre} (with $k=0$).
Moreover we also have for $N\geq 2$ 
\begin{align}
Z_{\Lambda}^{(n)}\left[J; 1, \ldots , n \right]  &=
 Z_{\Lambda}^{-1} \dfrac{\delta^{n} Z_{\Lambda} }{\delta J(1), \ldots  \delta J(n)} \nonumber \\
&=\left\langle   Z_{R}^{(n)}\left[J + \varphi; 1, \ldots , n \right]      \right\rangle \, ,
\end{align}
which is not a very useful result except for the case $n=1$ which gives us the exact relation
$$
\Phi_{\Lambda}\left[ J; 1 \right] =\left\langle\Phi_{R}\left[ J+\varphi; 2 \right] \right\rangle =
R^{\Lambda}_{0}(1,2) \left\langle \varphi (2) \right\rangle  \, ,
$$
from which we can guess the MF equation 
\begin{equation}\label{MF1}
\varphi_{MF} (1)=P^{\Lambda}_{0}(1,2) \Phi_{R} \left[ J+\varphi_{MF}; 2 \right] \, ,
\end{equation}
which we derive again now on more solid grounds.

The MF approximation is defined as usual as
\begin{subequations}
\begin{align}
 Z_{\Lambda, \, MF}&=\exp \left( -\mathcal{H}_{J}\left[ J,\varphi_{MF}\right]\right) \, , \\
\left. \dfrac{\delta \mathcal{H}}{\delta \varphi_{MF} }\right\vert_{J}&=0 \, . \label{oo}
\end{align}
\end{subequations}
Clearly the stationarity condition\ \eqref{oo} coincides with equation\ \eqref{MF1}. A short calculation
will show that the MF Gibbs
free energy is given by \cite{Cai-Mol}
\begin{equation}\label{GMF1}
 \Gamma_{\Lambda , \, MF}\left[ \Phi \right] =  \Gamma_{R}\left[ \Phi \right] -\dfrac{1}{2}
\Phi \cdot P^{\Lambda}_{0}   \cdot \Phi \, .
\end{equation}
The $2$-point vertex function and its inverse are then easily derived
 from\ \eqref{GMF1} 
\begin{align}
\Gamma_{\Lambda, \,MF}^{(2)}&= \Gamma_{R}^{(2)} -  P^{\Lambda}_{0}  \nonumber \, , \\
W_{\Lambda, \, MF}^{(2)}&=\left( 1 -W_{R}^{(2)} \cdot P^{\Lambda}_{0}  \right)^{-1}\cdot W_{R}^{(2)}  \nonumber \, .
\end{align}
To see that  $\Gamma_{\Lambda, \, MF}\left[ \Phi \right]$ is a rigorous upper bound to $\Gamma_{\Lambda }\left[ \Phi \right]$
we rewrite
\begin{equation}
Z_{\Lambda }\left[ J \right] = \left\langle \exp W_{R }\left[ J + \varphi \right] \right\rangle_{P^{\Lambda}_{0}} \, ,
\end{equation}
where the brackets denote a Gaussian average (see appendix~A). Applying Young inequalities\ \eqref{Young}
yields
\begin{align}
Z_{\Lambda }\left[ J \right] & \geq  
\left\langle \exp \left( \left( J+ \varphi \right) \cdot \Phi - \Gamma_{R}\left[\Phi \right]  \right) 
 \right\rangle_{P^{\Lambda}_{0}} \nonumber \qquad & \forall J, \, \forall \Phi \, , \\
& \geq \exp \left( - \Gamma_{R}\left[\Phi \right] +J \cdot \Phi \right) 
\left\langle  \exp\left( \Phi \cdot \varphi \right) \right\rangle_{P^{\Lambda}_{0}}  \qquad & \forall J, \, \forall \Phi \, , \nonumber \\
& \geq \exp \left( - \Gamma_{R}\left[\Phi \right] +J \cdot \Phi  + \dfrac{1}{2}
 \Phi \cdot P^{\Lambda}_{0} \Phi\right)  \qquad & \forall J, \, \forall \Phi \, .
\end{align}
Taking the log and making use of \eqref{GMF1}
$$
W_{\Lambda }\left[ J \right] \geq - \Gamma_{\Lambda, \,MF}\left[\Phi \right] +J \cdot \Phi  \qquad \forall J, \, \forall \Phi \, ,
$$
and therefore, for all $\Phi$
$$
\Gamma_{\Lambda,\, MF}\left[\Phi \right] \geq \sup_{J}\left\lbrace J \cdot \Phi -W_{\Lambda }\left[ J \right] \right\rbrace
\equiv \Gamma_{\Lambda}\left[\Phi \right] \, ,
$$
QED. It is obvious that we can extend these result to the $k$-systems and therefore one has, for any $k$
\begin{equation}
 \overline{\Gamma}^{\Lambda}_{k, \,MF} \left[ \Phi \right] =  \Gamma_{R}\left[ \Phi \right] -\dfrac{1}{2}
\Phi \cdot P^{\Lambda}_{k}   \cdot \Phi \geq  \overline{\Gamma}^{\Lambda}_{k } \left[ \Phi \right]\, ,
\end{equation}
from which it will be easy for the reader  to deduce that for all $0\leq k \leq \Lambda$ one has the rigorous
bound :
\begin{equation}\label{GMFk}
\Gamma^{\Lambda}_{k } \left[ \Phi \right]\leq \Gamma_{\Lambda,\, MF}\left[\Phi \right]  \qquad  \forall \Phi \, .
\end{equation}

\end{document}